\documentclass[manuscript]{acmart}

\AtBeginDocument{%
  \providecommand\BibTeX{{%
    \normalfont B\kern-0.5em{\scshape i\kern-0.25em b}\kern-0.8em\TeX}}}

\setcopyright{acmcopyright}
\copyrightyear{2018}
\acmYear{2018}
\acmDOI{XXXXXXX.XXXXXXX}
\usepackage{algorithm}
\usepackage{listings, xcolor}

\definecolor{verylightgray}{rgb}{.97,.97,.97}

\lstdefinelanguage{Solidity}{
	keywords=[1]{anonymous, assembly, assert, balance, break, call, callcode, case, catch, class, constant, continue, constructor, contract, debugger, default, delegatecall, delete, do, else, emit, event, experimental, export, external, false, finally, for, function, gas, if, implements, import, in, indexed, instanceof, interface, internal, is, length, library, log0, log1, log2, log3, log4, memory, modifier, new, payable, pragma, private, protected, public, pure, push, require, return, returns, revert, selfdestruct, send, solidity, storage, struct, suicide, selfdestruct, super, switch, then, this, throw, transfer, true, try, typeof, using, value, view, while, with, addmod, ecrecover, keccak256, mulmod, ripemd160, sha256, sha3}, % generic keywords including crypto operations
	keywordstyle=[1]\color{blue}\bfseries,
	keywords=[2]{address, bool, byte, bytes, bytes1, bytes2, bytes3, bytes4, bytes5, bytes6, bytes7, bytes8, bytes9, bytes10, bytes11, bytes12, bytes13, bytes14, bytes15, bytes16, bytes17, bytes18, bytes19, bytes20, bytes21, bytes22, bytes23, bytes24, bytes25, bytes26, bytes27, bytes28, bytes29, bytes30, bytes31, bytes32, enum, int, int8, int16, int24, int32, int40, int48, int56, int64, int72, int80, int88, int96, int104, int112, int120, int128, int136, int144, int152, int160, int168, int176, int184, int192, int200, int208, int216, int224, int232, int240, int248, int256, mapping, string, uint, uint8, uint16, uint24, uint32, uint40, uint48, uint56, uint64, uint72, uint80, uint88, uint96, uint104, uint112, uint120, uint128, uint136, uint144, uint152, uint160, uint168, uint176, uint184, uint192, uint200, uint208, uint216, uint224, uint232, uint240, uint248, uint256, var, void, ether, finney, szabo, wei, days, hours, minutes, seconds, weeks, years},	% types; money and time units
	keywordstyle=[2]\color{teal}\bfseries,
	keywords=[3]{block, blockhash, coinbase, difficulty, gaslimit, number, timestamp, msg, data, gas, sender, sig, value, now, tx, gasprice, origin},	% environment variables
	keywordstyle=[3]\color{violet}\bfseries,
	identifierstyle=\color{black},
	sensitive=false,
	comment=[l]{//},
	morecomment=[s]{/*}{*/},
	commentstyle=\color{gray}\ttfamily,
	stringstyle=\color{red}\ttfamily,
	morestring=[b]',
	morestring=[b]"
}
\usepackage{balance}
\usepackage{subfig}

\begin{document}

\title{VOLCANO: Detecting Vulnerabilities of Ethereum Smart Contracts Using Code Clone Analysis}

\author{Noama Fatima Samreen}

\email{noama.samreen@ryerson.ca}
\affiliation{%
  \institution{Ryerson University}
  \streetaddress{Department of Computer Science}
  \city{Toronto}
  \state{Ontario}
  \country{Canada}
}

\author{Manar H. Alalfi}

\email{manar.alalfi@ryerson.ca}
\affiliation{%
  \institution{Ryerson University}
  \streetaddress{Department of Computer Science}
  \city{Toronto}
  \state{Ontario}
  \country{Canada}
}

\begin{abstract}Ethereum Smart Contracts based on Blockchain Technology (BT) enables monetary transactions among peers on a blockchain network independent of a central authorizing agency. Ethereum Smart Contracts are programs that are deployed as decentralized applications, having the building blocks of the blockchain consensus protocol. This enables consumers to make agreements in a transparent and conflict-free environment. However, there exists some security vulnerabilities within these smart contracts that are a potential threat to the applications and their consumers and have shown in the past to cause huge financial losses. This paper presents a framework and empirical analysis that use code clone detection techniques for identifying vulnerabilities and their variations in smart contracts. Our empirical analysis is conducted using Nicad code clone detection tool on a dataset of approximately 50k Ethereum smart contracts.  We evaluated VOLCANO on two datasets, one with confirmed vulnerabilities and another with approximately 50k random smart contracts collected from the Etherscan\cite{Etherscsan}. Our approach shows an improvement in detection of vulnerabilities in terms of coverage and efficiency when compared to two of the publicly available static analysers to detect vulnerabilities in smart contracts. To the best of our knowledge, this is the first study that uses a clone detection technique to identify vulnerabilities and their evolution in Ethereum smart contracts.
\end{abstract}




\maketitle

%\section*{Declarations}
%\subsection{Funding}
%This work is supported in part by the Natural Sciences and Engineering Research Council of Canada (NSERC).The funders had no role in study design, data collection and analysis of the experiment.
%\subsection{Conflicts of interest/Competing interests}
%The authors declare there are no competing interests.

%\subsection{Availability of data and material}
%The datasets used in this research and a catalog of all the vulnerabilities detected in the Evaluation Dataset by VOLCANO as well as SmartCheck and Slither are available on GitHub\cite{NoamaDataset} and can be made publicly available upon request. The data were derived from the following resources available in the public domains: Etherescan \cite{Etherscsan}, State of the Dapps \cite{SDApps} and Dapp Radar \cite{DAppRadar}.
%\subsection{Code availability}
%The code of the findings of this study are available on request from the corresponding author, [Noama Samreen]. 
%\subsection{Additional declarations for articles in life science journals that report the results of studies involving humans and/or animals}
%Not Applicable
%\subsection{Ethics approval}
%Not Applicable
%\subsection{Consent to participate}
%Not Applicable
%\subsection{Consent for publication}
%Not Applicable

\section{Introduction}\label{sec1}
The advent of Ethereum\cite{Ethereum} at the end of 2015 changed the way everyone looked at blockchain technology. Ethereum Smart Contracts leverages blockchain technology to enable transactions among peers on a blockchain network independent of a central authorizing agency. However, recent research to identify the existence of security vulnerabilities in Ethereum Smart Contracts have shown that many applications have been exposed to attacks because of vulnerabilities found in application level of Ethereum Smart Contracts. These vulnerabilities are a potential threat to their consumers and have caused immense financial losses in the past. In a previous study, we have conducted a literature survey to highlight the seriousness of these security vulnerabilities and compared some of the available tools that target the detection of these security vulnerabilities in Ethereum Smart Contracts \cite{surveyVul}. However, the revolutionary features of Ethereum and the continual version updates of Solidity Programming language makes detection of vulnerabilities and updating smart contracts much harder than traditional programs. Researchers have explored different ways to identify these vulnerabilities. Prior code clone detection research in Ethereum smart contracts have shown that there existed almost 81\% of code cloning in smart contracts\cite{hassan}. Therefore, we believe that if a vulnerable code pattern can be obtained from a code base of confirmed vulnerabilities, then we might be able to use the vulnerable code pattern as a code-signature to identify other vulnerable smart contracts. Hence, code clone detection technique could be an ideal technique for vulnerabilities detection and should provide accurate results.
Investigating code cloning in Ethereum smart contracts from a vulnerability detection perspective, we need to address the following research questions: 
\begin{itemize}
    \item RQ1: Can we use clone detection techniques to generate a pattern of vulnerable code extracted from a dataset of vulnerable/exploited smart contracts?
    \item RQ2: Can we use the vulnerability signature to detect similar vulnerabilities in a dataset of approximately 50k smart contracts?
    \item RQ3: Can we find variants of these vulnerabilities in a Solidity programming language version-sorted dataset of smart contracts?
\end{itemize}

To our knowledge, this is the first study that uses clone detection technique to identify the vulnerabilities and their evolution in Ethereum smart contracts. Moreover, this paper presents the following contributions while investigating the research questions:
\begin{enumerate}
    \item An empirical analysis of code clone detection technique to identify vulnerabilities in approximately 50k Ethereum smart contracts. 
    \item An comparison of the results produced by the proposed technique of detecting vulnerabilities when compared to two of the publicly available static analysers to detect vulnerabilities in smart contracts. 
    \item Vulnerability evolution analysis across different versions of the Solidity programming language;
    \item a literature review on available approaches of using code cloning to detect vulnerabilities in a system.
\end{enumerate} 
After the introduction, Section 2 briefly highlights the background of Ethereum smart contracts and code cloning in general. Section 3 presents an overview of our approach, VOLCANO and a description of the datasets used for this study; Section 4 shows the results of applying a cross-code clone detection tool to the datasets to detect vulnerabilities. Section 5 presents the evaluation of this approach when compared to two other publicly available static analysers to detect vulnerabilities in smart contracts. Section 6 describes the studied approaches in the literature review of vulnerability identification using code-clone detection technique, followed by a conclusion in Section 6.
%The analysis and results are discussed in Section 4. Section 5 concludes the paper
\section{Background}
\subsection{Smart Contracts}
Ethereum uses Smart-Contracts\cite{Ethereum}, which are computer programs that directly controls the flow or transfer of digital assets.  Implementing a Smart-Contract use case can pose few security challenges, like public visibility of the complete source code of an application on a network, and validation and verification of the source code. Moreover, the immutability of Blockchain technology makes patching discovered vulnerabilities in already deployed Smart-Contracts impossible.

\textbf{Solidity}

Smart contracts are typically written in a high-level Turing-complete programming language such as Solidity\cite{Solidity}, and then compiled to the Ethereum Virtual Machine (EVM) bytecode\cite{Ethereum}, a low-level stack-based language.

\subsection{Code Cloning}
Code clone Terminology used in this study\cite{nicad}:
\begin{enumerate}
    \item \textbf{Code Fragment:} It is a pattern of lines of code which can be of any granularity, e.g., function definition, or sequence of statements.

    \item \textbf{Code Clone:} A code fragment is known as a code clone if the code fragment is either identical or similar to another code fragment. 

    \item \textbf{Clone Pair:} A pair of code fragments that are identical or similar.

    \item \textbf{Clone Class:} More than two code fragments that are identical or similar.

    \item \textbf{Code Clone Type:} Depending of the percentage of similarity between two code fragments, Code clones are classified into following types: 
    \begin{enumerate} 
        \item \textbf{Code clones of Type-1:}  These are code fragments that are identical with variation in comments and blank spaces.
        \item \textbf{Code clones of Type-2:} The code fragments that are identical with variations in identifier names, literals, comments and blank spaces. These clones are also called blind renamed clones, where Blind Renaming is the process of renaming all the identifiers to \textbf{X} without keeping a record of previously renamed identifier. 
        \item \textbf{Code clones of Type-2c:} Similar to the Type 2 code clones, this type of code clones is consistently renamed, which is process of renaming the identifiers to \textbf{X}\textit{n}. Where, \textit{n} is the number of occurrences of the identifier with respect to the previously renamed identifier.
        \item \textbf{Code clones of Type 3-1:} These are code clones that have similar code fragments with further modifications than Type-2 code clones like changed, added, or removed statements, in addition to variations in identifiers, literals, types, white-space, layout and comments.
        \item \textbf{ Code clones of Type 3-2c:} These are code clones that have similar code fragments with further modifications than Type-2c code clones. Type-3-2c can be considered as a hybrid of the Type 3-1 and  Type 2c code clones in that they are similar code fragments with changed, added, or removed statements, in addition to variations in identifiers, literals, types that are  consistently renamed.
        \item \textbf{ Code clones of Type 4:} These are known as Functional Clones with two or more code fragments that perform the same computation but are implemented by different syntactic variants.
\end{enumerate}
\end{enumerate}

\section{Approach}
\begin{figure*}
\centerline{\includegraphics[width=1\textwidth]{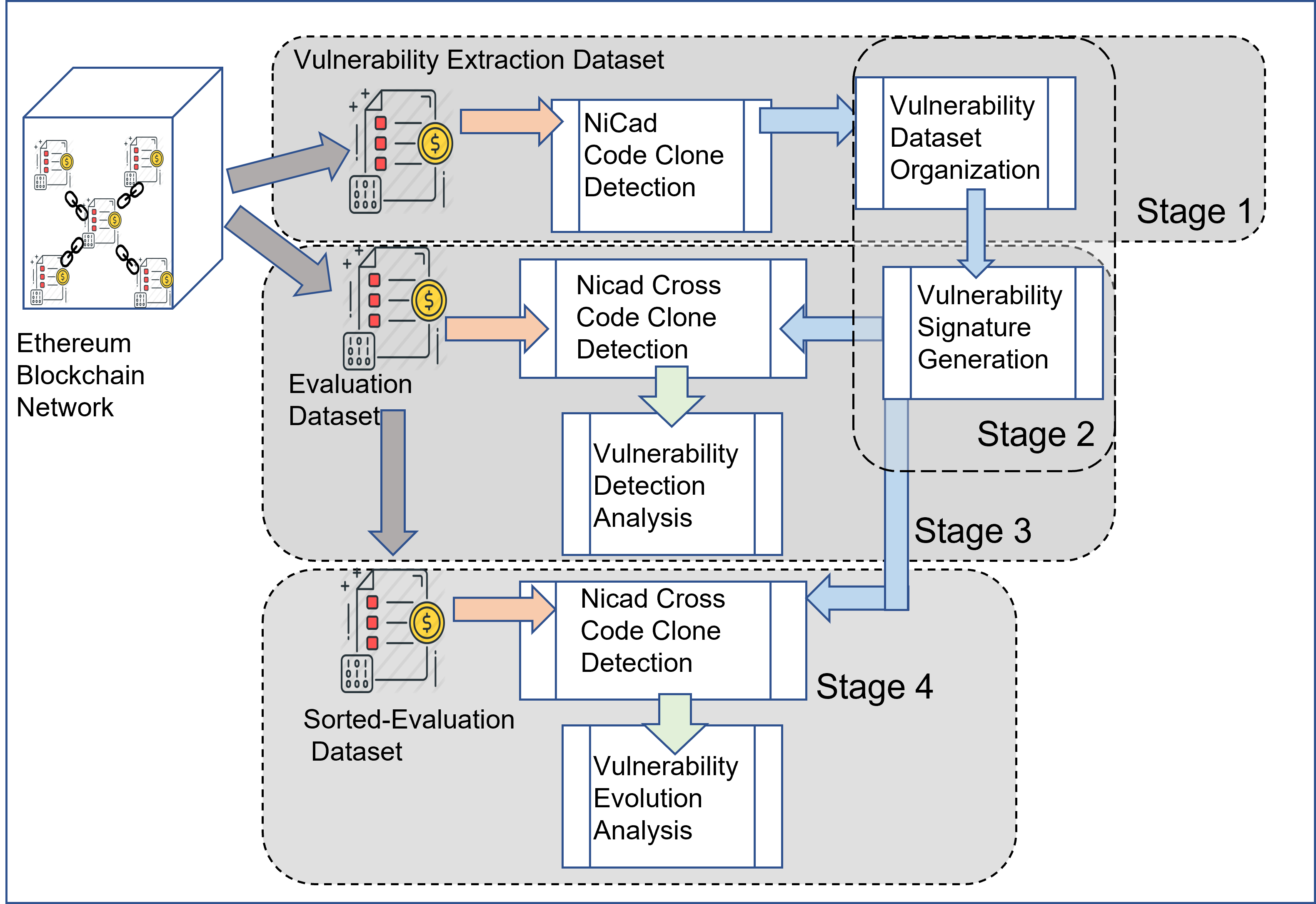}}
\caption{Approach}
\label{fig:crossApproach}
\end{figure*}

Current research in vulnerability detection in Ethereum smart contracts use static, dynamic or combined program analysis. We aim at developing a different approach to detect vulnerabilities in Ethereum smart contracts to achieve high recall and precision. We believe that the evolution of programming languages that supports development of Ethereum smart contracts is proportional to the evolution of known vulnerabilities in smart contracts. Therefore, we leverage the code clone detection technique based on static analysis of smart contracts in a 4-step process to identify and investigate the evolving nature of vulnerabilities in Ethereum smart contracts. 
[See Figure \ref{fig:crossApproach}]

Our approach, VOLCANO, consists of four main stages: 
\begin{enumerate}
    \item \textbf{Vulnerability Dataset organization}
    
    This dataset consists of the smart contracts that have confirmed vulnerabilities in them. The Vulnerability Dataset was collected from multiple open-source dataset repositories available on GitHub\cite{SmartBugs}. However, we found inconsistencies in the categorization of smart contract vulnerabilities in our dataset sources. Therefore, we performed code clone detection using Nicad tool on the Vulnerability Dataset to generate a vulnerability signature for each of the vulnerability type clone class present in the dataset and thereby efficiently segregate and categorize the vulnerabilities. The final organization of our Vulnerability Dataset is presented in the Table \ref{ExtractionDataset}. 
    For this empirical analysis, we set the Nicad tool to detect clones at the function granularity level. The main aim of this step is to cluster vulnerabilities into different classes of clones. These classes can be used to generate a vulnerability signature for each of the vulnerability clone class identified in this step and also to increase the efficiency of identifying vulnerability clones in the Evaluation Dataset by having vulnerabilities clone classes segregated in separate folders.
    The setup to segregate smart contracts in the Vulnerability Dataset followed the detection of type 3-2 and type 3-2c code clones with a maximum difference threshold ranging from 0\% to 30\% to extract and categorize clones of vulnerabilities. 
    The generated code clone classes were manually analyzed to confirm the existence of a vulnerability in them and categorize it. 

    \item \textbf{Vulnerability signature generation using code clone detection in the Vulnerability Dataset}
    
    Following the above step of segregating the code clone classes that contain different types of vulnerabilities, a vulnerability signature was derived for each of the clone classes detected. 
    The vulnerability signature is a code pattern that corresponds to a type of known vulnerabilities in Ethereum smart contracts with a maximum difference threshold of 30\%. This was to account for slight variations in the vulnerability signature pattern matching process. 
    
    \item \textbf{Vulnerability identification using cross-code clone detection between organized Vulnerability Dataset and Evaluation Dataset}
    
    Our second dataset, Evaluation Dataset, contains smart contracts that were randomly collected over a period of time from Etherscan\cite{Etherscsan}. We collected 49,969 Ethereum smart contracts written in Solidity programming language and having at least one transaction from Etherscan\cite{Etherscsan}. These smart contracts were not initially filtered for vulnerabilities and we used this dataset in this stage of our approach to identify vulnerabilities using the vulnerability signatures generated from the Vulnerability Dataset in the previous stage. 
    The setup to identify the vulnerability code clones in the Evaluation Dataset was to detect type 3-2 and type 3-2c code clones with a maximum difference threshold ranging from 0\% to 30\%. 
    To efficiently identify code clone classes with vulnerability signature, we added a comment before the start of a vulnerability signature to highlight the vulnerability type.  
    
    \item \textbf{Vulnerability evolution analysis across different versions of the Solidity programming language}
    
    The smart contracts in the Evaluation Dataset are organized using a python script that sorts the smart contracts based on their Solidity programming language version number.  Following this sorting, the cross-code clone detection is repeated on the Sorted-Evaluation Dataset to analyze the difference between a vulnerability clone class from a smart contract of an earlier version of Solidity programming language to that of the same in a newer version. 
\end{enumerate}
 
%\subsection{Smart contracts segregation and dataset organization}

%\subsection{Vulnerability signature Generation using Code Clone Detection}

\begin{table}
\caption{VULNERABILITY DATASET ORGANIZATION}
\begin{center}
 \begin{tabular}{|p{0.5cm}|p{4.5cm}|p{2cm}|p{4.5cm}|}
 \hline
\textbf{S.no}&\textbf{Vulnerability Type}& \textbf{No. of Smart Contracts} & \textbf{No. of Vulnerability Instances}\\
\hline
1&\textbf{Re-entrancy}& 	37  & 42\\

2&\textbf{Denial of Service (DOS)}&   17 & 22\\

3&\textbf{Integer Underflow/Overflow}& 16 & 20\\

4&\textbf{Call-To-Unknown}& 44 & 45\\

5&\textbf{Out-of-Gas exception}& 14 & 20\\

6&\textbf{Mishandled Exceptions}& 54 & 58\\

7&\textbf{Mismatched Typecasting}& 7 & 7\\

8&\textbf{Weak Modifiers}& 33 & 68\\
\hline
&\textbf{Total} & 222 & 282\\
\hline
\end{tabular}
\end{center}
\label{ExtractionDataset}
\end{table} 

%\subsection{Vulnerability identification in the Evaluation Dataset using Cross-Code Clone Detection}

%\subsection{Understanding the evolution of the vulnerabilities in Ethereum smart contracts}

\section{Results and Discussion}
This study conducts an empirical analysis to understand the state of vulnerabilities in Ethereum smart contracts by trying to study the patterns, extent, and variations of known vulnerabilities in Ethereum smart contracts by using code clone detection techniques. The research questions highlighted in this research study are answered below:

\subsection*{\textbf{RQ1: Can we use clone detection techniques to generate a pattern of vulnerable code extracted from a dataset of vulnerable/exploited smart contracts?}}

\textbf{\textit{Motivation:}} We have reviewed several research studies that have explored the possibility of applying code clone detection techniques to detect vulnerabilities in software systems [See section \ref{relatedwork}].  However, none of the research works we reviewed used code clone detection technique to form vulnerability signature of vulnerable code patterns from a dataset of confirmed vulnerabilities in Ethereum smart contracts and further detect these vulnerabilities in a generic dataset to understand the extent and evolution of these vulnerabilities. 

 \textbf{\textit{Findings:}}   
 Vulnerable code can form code clone classes. We use the Nicad\cite{nicad} code clone detection tool to extract these classes from a dataset of confirmed vulnerabilities in Ethereum smart contracts. 
 Table \ref{ExtractionDataset} shows the preliminary result of the vulnerability type classification and signature generation. The Table \ref{ExtractionDataset} is a snapshot of the vulnerable code clone classes at a 0\% difference threshold detected in our Vulnerability Dataset. We found eight vulnerability types in our dataset: \textit{Reentrancy, Denial of Service(DoS), Integer Underflow/Overflow, Call-to-Unknown, Out-of-Gas Exception, Mishandled Exceptions, Mismatched Typecasting, Weak Access Modifiers.} These vulnerability types had a varied number of smart contracts that belonged to multiple code clone classes. The vulnerable code clone classes were analyzed to derive the vulnerability signature for each vulnerability. We validated the vulnerability signatures generated by comparing them to the code examples from the research available in this field \cite{surveyVul}.
Each of the vulnerability types is described next with references to the vulnerability signatures derived for each type. 

\subsubsection*{Call-to-Unknown Vulnerability}
Listings \ref{ctu1},\ref{ctu2} presents the vulnerability-signatures extracted for the \textit{Call-to-Unknown} vulnerability. This vulnerability is related to access control issues. Accessing a contract's functionality through its public or external functions with improper function signatures or insecure visibility settings gives rise to this vulnerability in a smart contract.
In listing \ref{ctu1}, a vulnerable smart contract gets initialized by an address which gets decided by line 2. This is one of the common coding practices for defining access permissions like withdrawing the contract's funds etc. This makes the smart contract vulnerable as the initialization function can be called multiple times and by anyone via \textit{msg.sender}. 
Listing \ref{ctu2} is a code pattern that uses the built-in low level function \textit{delegatecall()}. When a contract executes \textit{delegatecall()} to call another contract, then the latter contract's code gets executed. The call-to-unknown vulnerability rises when the contract's interface is not checked before calling it or the return value of this external call is not checked. 

\begin{lstlisting}[language=Solidity, label={ctu1}, caption={Call-to-unknown Vulnerability Signature-1}, frame=trBL]
 
function initialize() public {
	new_owner = msg.sender;
}
\end{lstlisting}

\begin{lstlisting}[language=Solidity, label={ctu2}, caption={Call-to-unknown Vulnerability Signature-2 },frame=trBL] 
function() payable {
    if (msg.data.length > 0)
      owner.delegatecall(msg.data); 
  }
\end{lstlisting}

\subsubsection*{Denial of Service (DoS) Vulnerability}
Listings \ref{dos1},\ref{dos2},\ref{dos3},\ref{dos4} presents the extracted vulnerability-signatures for the \textit{Denial of Service (DoS)} vulnerability. An exploitation of the DoS vulnerability makes a smart contracts unavailable for service on the Ethereum network. The major scenarios that can make a smart contract vulnerable to DoS attack are highlighted in listings \ref{dos1}, \ref{dos2}, \ref{dos3}, \ref{dos4}. 
Listings \ref{dos1} and \ref{dos2} shows the usage of built-in functions \textit{suicide()} and \textit{selfdestruct()} without checking the ownership of the smart contract or access permissions. Any external user can gain control of the smart contract and call this function to suspend all the transactions of the smart contract and make it unavailable for service. 
Listing \ref{dos3} shows the usage of the external call function \textit{send()} inside a \textit{for-statement}. If the \textit{send()} function fails to execute, then the transaction is reverted and the loop is halted and make the smart contract unavailable for service.
Similar to listing \ref{dos3}, listing \ref{dos4} shows the usage of the built-infunction \textit{require()} inside a \textit{for-statement}. The \textit{require()} function takes a condition that cannot be detected until execution time. This condition can be dependent on variables such as inputs or return values from calls to external contracts. Therefore, using this function inside a \textit{for-statement} can halt the loop and make the smart contract unavailable for service.

\begin{lstlisting}[language=Solidity, label={dos1}, caption={DoS Vulnerability Signature-1},frame=trBL]
function kill(address malicious) external {
    suicide(malicious);
    }
    \end{lstlisting}

\begin{lstlisting}[language=Solidity, label={dos2}, caption={DoS Vulnerability Signature-2},frame=trBL] 
function kill(address malicious) external {
    selfdestruct(malicious);
    }
    \end{lstlisting}

\begin{lstlisting}[language=Solidity, label={dos3}, caption={DoS Vulnerability Signature-3},frame=trBL] 
function sendPayments() public returns (bool){
         for(uint i=0;i<n;i++) {
            addresses.send(msg.sender);
        }    return true;
    }
 \end{lstlisting}

\begin{lstlisting}[language=Solidity, label={dos4}, caption={DoS Vulnerability Signature-4},frame=trBL] 
function sendPayments() public returns (bool){
         for(uint i=0;i<n;i++) {
             require(addresses.send(msg.sender));
        }    
        return true;
    }
 \end{lstlisting}

\subsubsection*{Reentrancy Vulnerability}
Listing \ref{reentrancy} shows the vulnerability signature derived for the re-entrancy vulnerability. A re-entrancy attack can drain a smart contract of all its ether, which was first discovered in the Decentralised Autonomous Organization (DAO) attack of 2016. As the name of the vulnerability suggests, re-entrancy occurs when an external contract makes multiple calls to a contract before the initial call’s execution is completed. In listing \ref{reentrancy}, the vulnerability scenario arises when an external function call \textit{send()} is called to send out ether before updating the state variables. This causes the fallback function of the contract receiving the ether is executed. If the fallback function of the receiving contract calls the withdraw of ether again, then the \textit{send()} is executed again as the state variable that hold the total balance of ether is updated after the execution of the \textit{send()} function is completed. 

\begin{lstlisting}[language=Solidity, label={reentrancy}, caption={Re-entrancy Vulnerability Signature},frame=trBL] 
function externalSend(uint amountToSend) {
	if(balance >= amountToSend)
	msg.sender.call.value(amountToSend)();
	balance -= amountToSend; \\state variable updated after external call function is executed
}
 \end{lstlisting}

\subsubsection*{Integer Underflow/Overflow Vulnerability}
Another common vulnerability in Ethereum smart contracts is Integer underflow and overflow error. This vulnerability is not specific to Ethereum smart contracts construct; However, these make the smart contracts exposed to malicious attacks with the motive of draining the victim smart contract of its ether. Listing \ref{integer} shows the vulnerability signature of this type which includes a simple function that does not check for integer underflow or overflow before performing arithmetic operations. 

\begin{lstlisting}[language=Solidity, label={integer}, caption={Integer Underflow/Overflow Vulnerability Signature},frame=trBL] 
function externalSend(uint amountToSend) {
	if(balance >= amountToSend)
	msg.sender.call.value(amountToSend)();
	balance -= amountToSend; \\
}
 \end{lstlisting}

\subsubsection*{Mishandled Exceptions Vulnerability}
The exception handling in Solidity programming language is not even for the exceptions thrown during external calls like \textit{send(), call(), delegatecall()}. The exceptions do not propagate in the code. These external calls return a Boolean value, which will be set to false when an exception is thrown. However, when this Boolean value is unchecked, the programmers are not notified in any other way about the exception and can expose smart contracts to malicious transactions that may result in the exploitation of this vulnerability. Listing \ref{misEx} shows the vulnerability signature for this type. 
\begin{lstlisting}[language=Solidity, label={misEx}, caption={Mishandled Exceptions Vulnerability Signature},frame=trBL] 
function externalCall(uint str) {
	msg.sender.delegateCall(str); \\without checking for return value
}
 \end{lstlisting}

\subsubsection*{Weak Access Modifiers Vulnerability}
Listing \ref{weak} shows the vulnerability signature for \textit{weak  access modifiers} vulnerability. When a modifier for a function or a parameter is not explicitly specified, then the default access modifier is \textit{public}. The \textit{public} access modifier makes the smart contract vulnerable to the \textit{weak access modifiers} vulnerability as these can be manipulated by observing the hash values of the blocks appended to the Ethereum network.

\begin{lstlisting}[language=Solidity, label={weak}, caption={Weak Access Modifiers Vulnerability Signature},frame=trBL] 
function initialize() public { //weak access modifier for the function initialize
	new_owner = msg.sender;
}
 \end{lstlisting}

\begin{lstlisting}[language=Solidity, label={outofGasEx}, caption={Out-of-Gas Exception Vulnerability Signature},frame=trBL] 
function externalSend(uint amountToSend) {
	if(balance >= amountToSend)
	 msg.sender.send(amountToSend); \\gasless-send
}
 \end{lstlisting}
 
\subsubsection*{Out-of-Gas Exception Vulnerability}
Listing \ref{outofGasEx} shows the vulnerability signatures derived from the code clone classes for \textit{Out-of-Gas Exception} vulnerability in the Vulnerability Dataset. When \textit{send()} external function call is used to ether to another smart contract, the fallback function of the receiving smart contract will be executed. If this function contains code that would cost more than 2300 gas units to get executed, then the fallback function will revert due to insufficient gas. The smart contract that was supposed to send the ether will keep the ether even though it was supposed to send it to the receiving smart contract because it is not notified of the exception raised due to an unexpected revert sequence. 

%\begin{Pattern}
%\begin{lstlisting}[language=Solidity, label={dos1}, caption={DoS Vulnerability Signature-1}] 
%function withdraw() {
%    if (msg.sender != 0) {
%      require(msg.sender.send(msg.value));
%      }
%  }
%     \end{lstlisting}
%\end{Pattern}
\subsection*{\textbf{RQ2: Can we use the vulnerability signature to detect similar vulnerabilities in a dataset of approximately 50k smart contracts?}}

\textbf{\textit{Motivation:}}

Code clone detection can identify code fragments that contains similar code patterns. Therefore, the cross-systems code clone detection feature provided by the Nicad tool was used in efficiently detecting similar vulnerabilities in the Evaluation Dataset.  

 \textbf{\textit{Findings:}} 
 
 Our analysis resulted in code clone classes consisting of vulnerable code fragments from the Vulnerability Dataset as well as code fragments from the smart contracts in the Evaluation Dataset. Therefore, we can confirm that the cross code clone detection technique could successfully identify vulnerability code patterns that are similar or identical to the vulnerability signatures generated. 
%\textbf{Vulnerability-Signatures Extraction}
%\textbf{Vulnerability Identification in Evaluation Dataset}
\begin{table}
\caption{CODE CLONES FROM A VULNERABILITY DATASET FOUND IN AN EVALUATION DATASET OF 50k SMART CONTRACTS COLLECTED}
\begin{center}
 \begin{tabular}{|p{2.5cm}|p{4cm}|p{4cm}|}
 \hline
& \textbf{Filtered, \newline Blind-Renamed}& \textbf{Filtered, \newline Consistent-Renamed}\\
\hline
\textbf{Clone Type} &\textbf{Type 3-2}\newline(Inclusive of Type 1, 2)& \textbf{Type 3-2c}\newline(Inclusive of Type 1, 2c)\\
\hline
\textbf{Clone Pairs}& 32034	& 19129 \\
\hline
\textbf{Total Clone Classes}& 212&  165  \\
\hline
\textbf{Max Difference Threshold}& 0\%& 30\% \\
\hline
\end{tabular}
\end{center}
\label{CloneDets}
\end{table}
Table \ref{CloneDets} shows the summary of the code clone classes discovered in the empirical analysis using cross-code clone detection between our Vulnerability Dataset and Evaluation Dataset. This resulted in 212 clone classes that were of Type 3-2 (Filtered and Blind-Renamed) with maximum difference threshold of 0\% and 165 clone classes of Type 3-2c (Filtered and Consistent-Renamed) with maximum difference threshold of 30\%. Table \ref{EvaluationDataset} shows the detection of vulnerability clones in the Evaluation Dataset. 
\begin{table}
\caption{VULNERABILITY CLONES IN THE EVALUATION DATASET}
\begin{center}
 \begin{tabular}{|p{4.5cm}|p{1.75cm}|p{1.75cm}|p{3cm}|}
 \hline
\textbf{Vulnerability Type}& \multicolumn{2}{|c|}{\textbf{Clone Classes}} & \textbf{Total Vulnerability Instances} \\
\hline
& \textbf{Type 3-2} & \textbf{Type 3-2c}&\\
\hline
\textbf{Re-entrancy}& 	28  & 22 & 10764\\

\textbf{Denial of Service (DOS)}&  23 &22 & 14189 \\

\textbf{Integer Underflow/Overflow}& 29 &23& 18892 \\

\textbf{Call-To-Unknown}& 32 &29 & 1078\\

\textbf{Out-of-Gas exception}& 17&14& 1552\\

\textbf{Mishandled Exceptions}& 14 & 11& 210\\

\textbf{Mismatched Typecasting}& 12& 8 & 268\\

\textbf{Weak Modifiers}& 18&15& 991\\
\hline
\textbf{Total} & 164 & 145 &  47944\\
\hline
\end{tabular}
\end{center}
\label{EvaluationDataset}
\end{table}
\subsection*{\textbf{RQ3: Can we find variants of these vulnerabilities in a Solidity programming language version-sorted dataset of smart contracts?}}

\textbf{\textit{Motivation:}}

As the Solidity programming language updates its versions to safeguard against the known vulnerabilities, the vulnerable code patterns change over time as well. This evolution can be studied by observing the variations in a vulnerability across different versions of Solidity programming language.

\textbf{\textit{Findings:}}  
%\textbf{Vulnerability Evolution in Sorted-Evaluation Dataset}
\begin{table}
\caption{EVALUATION DATASET ORGANIZATION BY SOLIDITY VERSION}
\begin{center}
 \begin{tabular}{|p{4.5cm}|p{2cm}|}
 \hline
\textbf{Solidity Version}& \textbf{No. of Smart Contracts}\\
\hline
\textbf{\^0.3}& 105	 \\
\textbf{\^0.4}& 28920	 \\
\textbf{\^0.5}& 13775	 \\
\textbf{\^0.6}& 4400	 \\
\textbf{\^0.7}& 2472	 \\
\textbf{\^0.8}& 297	 \\
\hline
\textbf{Total} & 49969 \\
\hline
\end{tabular}
\end{center}
\label{SortedEvaluationDataset}
\end{table} 

\begin{table}
\caption{VULNERABILITY CLONE CLASSES IN THE SORTED-EVALUATION DATASET BASED ON THE VERSIONS OF SOLIDITY}
\begin{center}
 \begin{tabular}{|p{2.75cm}|p{0.3cm}|p{0.3cm}|p{0.3cm}|p{0.3cm}|p{0.3cm}|p{0.3cm}|p{0.3cm}|p{0.3cm}|p{0.3cm}|p{0.3cm}|p{0.3cm}|p{0.3cm}|}
 \hline
\textbf{Vulnerability Type}& \multicolumn{2}{|c|}{\textbf{0.3}}& \multicolumn{2}{|c|}{\textbf{0.4}} & \multicolumn{2}{|c|}{\textbf{0.5}}& \multicolumn{2}{|c|}{\textbf{0.6}}& \multicolumn{2}{|c|}{\textbf{0.7}}& \multicolumn{2}{|c|}{\textbf{0.8}}\\ 
\hline
&\rotatebox[origin=c]{90}{\textbf{Type 3-2}} & \rotatebox[origin=c]{90}{\textbf{Type 3-2c}}&\rotatebox[origin=c]{90}{\textbf{Type 3-2}} & \rotatebox[origin=c]{90}{\textbf{Type 3-2c}}&\rotatebox[origin=c]{90}{\textbf{Type 3-2}} & \rotatebox[origin=c]{90}{\textbf{Type 3-2c}}&\rotatebox[origin=c]{90}{\textbf{Type 3-2}} & \rotatebox[origin=c]{90}{\textbf{Type 3-2c}}&\rotatebox[origin=c]{90}{\textbf{Type 3-2}} & \rotatebox[origin=c]{90}{\textbf{Type 3-2c}}&\rotatebox[origin=c]{90}{\textbf{Type 3-2}} & \rotatebox[origin=c]{90}{\textbf{Type 3-2c}}\\
\hline
\textbf{Re-entrancy}& 0 & 0 & 15 & 14 & 6 & 5 & 3 & 1 & 2 & 1 & 2 & 1 \\

\textbf{DoS}& 0 & 1 & 7 & 6 & 6 & 6 & 5 & 5 & 3 & 3 & 1 & 2\\

\textbf{Integer U/O}& 1 & 1 & 9 & 10 & 5 & 5 & 4 & 4 & 3 & 2 & 1 & 1\\

\textbf{Call To Unknown}& 2 & 2 & 14 & 14 & 6 & 5 & 3 & 4 & 4 & 3 & 1 & 1 \\

\textbf{Gas exception}& 0 & 0 & 11 & 8 & 3 & 3 & 1 & 1 & 1 & 1 & 1 & 1 \\

\textbf{Mis-Exceptions}& 0 & 0 & 8 & 5 & 3 & 2 & 1 & 2 & 1 & 1 & 1 & 1\\

\textbf{Mis-Typecasting}& 0 & 0 & 5 & 3 & 3 & 2 & 2 & 1 & 1 & 1& 1 &1\\

\textbf{Weak Modifiers}& 2 & 1 & 6 & 6 & 5 & 4 & 2 & 3 & 2 & 1 & 1 & 0\\
\hline
\textbf{Total}& 5 & 5 & 75 & 66 & 37 & 32 & 21 & 21 & 17 & 13 & 9 & 8\\
\hline
\end{tabular}
\end{center}
\label{EvaluationDatasetOrganized}
\end{table}

Table \ref{SortedEvaluationDataset} shows the sorted evaluation dataset based on their Solidity version number. The Sorted-Evaluation Dataset resulted in organization of the dataset into 6 sub-folders representing versions of Solidity programming language.  The smart contracts in our dataset used Solidity programming language with version ranging from \textit{0.3 to 0.8} (including the sub-versions for each version). 
Table \ref{EvaluationDatasetOrganized} shows the detection of vulnerability clones in the Sorted-Evaluation Dataset based on their Solidity version number. From the total of 164 vulnerability-related clone classes of Type 3-2 and 145 vulnerability clone classes of Type 3-2c, we analyse the distribution of clone classes across different versions of the Solidity programming language. 

\begin{table}
\caption{LOWEST SIMILARITY PERCENTAGE OF VULNERABILITY CLONE CLASSES IN THE SORTED-EVALUATION DATASET BASED ON THE VERSIONS OF SOLIDITY}
\begin{center}
 \begin{tabular}{|p{3cm}|p{1cm}|p{1cm}|p{1cm}|p{1cm}|p{1cm}|p{1cm}|}
 \hline
    & \multicolumn{6}{|c|}{\textbf{Type 3-2c}}\\ 
\hline
\textbf{Vulnerability Type}& \textbf{0.3}&\textbf{0.4}&\textbf{0.5}&\textbf{0.6}& \textbf{0.7}& \textbf{0.8}\\ 
\hline
\textbf{Re-entrancy}& NA & 98\% & 98\% & 96\% & 82\% & 70\% \\

\textbf{DoS}& 98\% & 98\% & 94\% & 92\% & 88\% & 82\% \\

\textbf{Integer U/O}& 88\% & 98\% & 98\% & 88\% & 76\% & 70\% \\

\textbf{Call To Unknown}& 92\% & 80\% & 82\% & 70\% & 70\%&70\% \\

\textbf{Gas exception}& NA & 98\% & 98\% & 82\% &78\%& 70\% \\

\textbf{Mis-Exceptions}& NA & 96\%& 88\% & 82\% & 82\% & 70\% \\

\textbf{Mis-Typecasting}& NA & 80\% & 80\% & 80\% & 70\% & 70\%  \\

\textbf{Weak Modifiers}& 98\% & 98\% & 98\% & 90\% & 82\% & 82\% \\
\hline
\end{tabular}
\end{center}
\label{SortedMaxDiff}
\end{table}

To analyze the changing patterns of vulnerabilities in newer Solidity versions, we observe the maximum difference threshold, in other words, lowest similarity percentage between vulnerability signature and smart contract's code fragments recorded for each of the vulnerability types in different versions of Solidity programming language. This is analysed only for the Type 3-2c code clone classes, as this type allows for a maximum difference threshold ranging from 0\% to 30\%.
In Table \ref{SortedMaxDiff}, we see that the similarity percentage decreases for newer versions of Solidity when compared to the earlier versions. It can be inferred from this observation that the vulnerability patterns undergo changes with updates to the Solidity programming language. 
\section{Evaluation}
We conducted a preliminary survey of available tools and technologies implemented to detect vulnerabilities in Ethereum smart contracts\cite{noamasurvey}. There are many static analysers in the market currently that provide vulnerabilities detection for Ethereum smart contracts. However, many of the available tools does not meet our requirements for evaluation. According to our research \cite{noamasurvey}, \cite{Monteiro2019ASO}, SmartCheck\cite{SmartCheck} and Slither\cite{Slither} static analysis tools conform to the following inclusion criteria relevant to our requirements:
\begin{enumerate}
    \item Publicly available with a Command Line Interface (CLI) - This requirement is suited to analyse our dataset of approx. 50k smart contracts. 
    \item Analyses smart contracts source code - The input to the tool is a smart contract \textit{.sol} file and this source code is only required to run the analysis. 
    \item Vulnerabilities Detector - The tool detects vulnerabilities or common bad coding practices in smart contracts.
\end{enumerate}

\begin{table}
\caption{VULNERABILITIES IDENTIFIED BY EACH TOOL PER CATEGORY IN VULNERABILITY DATASET}
\begin{center}
 \begin{tabular}{|p{4.5cm}|p{1.75cm}|p{1.75cm}|p{1.75cm}|}
 \hline
\textbf{Vulnerability Type}& \multicolumn{3}{|c|}{\textbf{Analysis Results}} \\
\hline
& \textbf{Slither} & \textbf{SmartCheck} & \textbf{VOLCANO}\\
\hline
\textbf{Re-entrancy}& 42 & 42 & 42\\

\textbf{Denial of Service (DOS)}&  0 & 0 & 22 \\

\textbf{Integer Underflow/Overflow}& 0 & 18 & 20 \\

\textbf{Call-To-Unknown}& 35 & 40 & 45\\

\textbf{Out-of-Gas exception}& 19 & 19 & 20 \\

\textbf{Mishandled Exceptions}& 50 & 50 & 58\\

\textbf{Mismatched Typecasting}& 0 & 7 & 7 \\

\textbf{Weak Modifiers}&  53 & 0  & 68\\
\hline
\textbf{Total} & 199 & 176 & 282 \\
\hline
\end{tabular}
\end{center}
\label{EvaluationVULDataset}
\end{table}

\begin{table}
\caption{VULNERABILITIES IDENTIFIED BY EACH TOOL PER CATEGORY IN EVALUATION DATASET}
\begin{center}
 \begin{tabular}{|p{4.5cm}|p{1.75cm}|p{1.75cm}|p{1.75cm}|}
 \hline
\textbf{Vulnerability Type}& \multicolumn{3}{|c|}{\textbf{Analysis Results}} \\
\hline
& \textbf{Slither} & \textbf{SmartCheck} & \textbf{VOLCANO}\\
\hline
\textbf{Re-entrancy}& 10764 & 8745 & 10764\\

\textbf{Denial of Service (DOS)}& 0  &  0 & 14189 \\

\textbf{Integer Underflow/Overflow}& 0 & 18725 & 18892\\

\textbf{Call-To-Unknown}& 1078 & 867 & 1078\\

\textbf{Out-of-Gas exception}&  1552 & 501 & 1552 \\

\textbf{Mishandled Exceptions}& 210 & 443 & 210\\

\textbf{Mismatched Typecasting}& 0 & 145 & 268 \\

\textbf{Weak Modifiers}& 391 &  0 & 991 \\
\hline
\textbf{Total} & 13995 & 29426 & 47944  \\
\hline
\end{tabular}
\end{center}
\label{EvaluationEVALDataset}
\end{table}
\subsection{Vulnerabilities Detected by the Analysis Tools}
We subjected both the datasets used in this study to Slither\cite{Slither} and SmartCheck\cite{SmartCheck} static analyser using their CLI. 
We observed that the number of vulnerability instances detected by both these tools is comparable to that detected by our approach with a few exceptions arising due to a couple of vulnerability types identified by our approach. For example, the Denial of service vulnerability type is not supported by either of the two tools used to evaluate our approach. Again, the Integer underflow and overflow vulnerability type was not detected by Slither and the Weak access modifiers type of vulnerabilities were missed by the SmartCheck analysis tool. 
Therefore, the results for the Vulnerability Dataset show a total of 199 vulnerability instances identified by Slither, 176 vulnerability instances identified by SmartCheck and 282 by our approach, VOLCANO. Also, in the table \ref{EvaluationEVALDataset}, we can observe the difference in number of vulnerability instances detected between VOLCANO and the two other tools because of the vulnerability pattern coverage implemented by our approach that accommodates for a maximum difference threshold of 30\%.

\subsection{Execution Time of the Analysis Tools}
To compare the time of the execution for each of the tools used for evaluation, we recorded each individual analysis time for both the datasets. This analysis time is measured by calculating the difference between the starting time and the ending time of each analysis.
Table \ref{AnalysisTimeRes} shows the average and total analysis times used by each tool to analyse 50k smart contracts in our Evaluation Dataset. In the table \ref{AnalysisTimeRes}, we can observe the significant difference in total analysis time between VOLCANO and the two other tools. This difference in analysis time between the three tools is because of the technique that each tool uses to analyse each smart contract. Slither and SmartCheck both parse the AST of the contract to identify vulnerabilities, whereas VOLCANO is based on signature matching using Nicad code clone detection tool. NiCad's incremental mode, in which a system that has already been analyzed is re-analyzed only for new clones, is responsible for significantly reducing the analysis time used by our approach. 

\begin{table}
\caption{ANALYSIS TIME RECORDED FOR EACH TOOL FOR EVALUATION DATASET}
\begin{center}
 \begin{tabular}{|p{4cm}|p{4.75cm}|p{4.75cm}|}
 \hline
\textbf{Analysis Tool}& \textbf{Average Analysis Time}& \textbf{Total Analysis Time} \\
\hline
\textbf{Slither} & 00:00:05 & 2 days, 21:26:38  \\
\textbf{SmartCheck} & 00:00:10 & 5 days, 18:43:12  \\
\textbf{VOLCANO}& 00:00:01 & 13:53:16 \\
\hline
\end{tabular}
\end{center}
\label{AnalysisTimeRes}
\end{table}

The datasets used in this research and a catalog of all the vulnerabilities detected in the Evaluation Dataset by VOLCANO as well as SmartCheck and Slither are available on GitHub\cite{NoamaDataset} and can be made publicly available upon request. 
\begin{figure*}
\centerline{\includegraphics[width=1\textwidth]{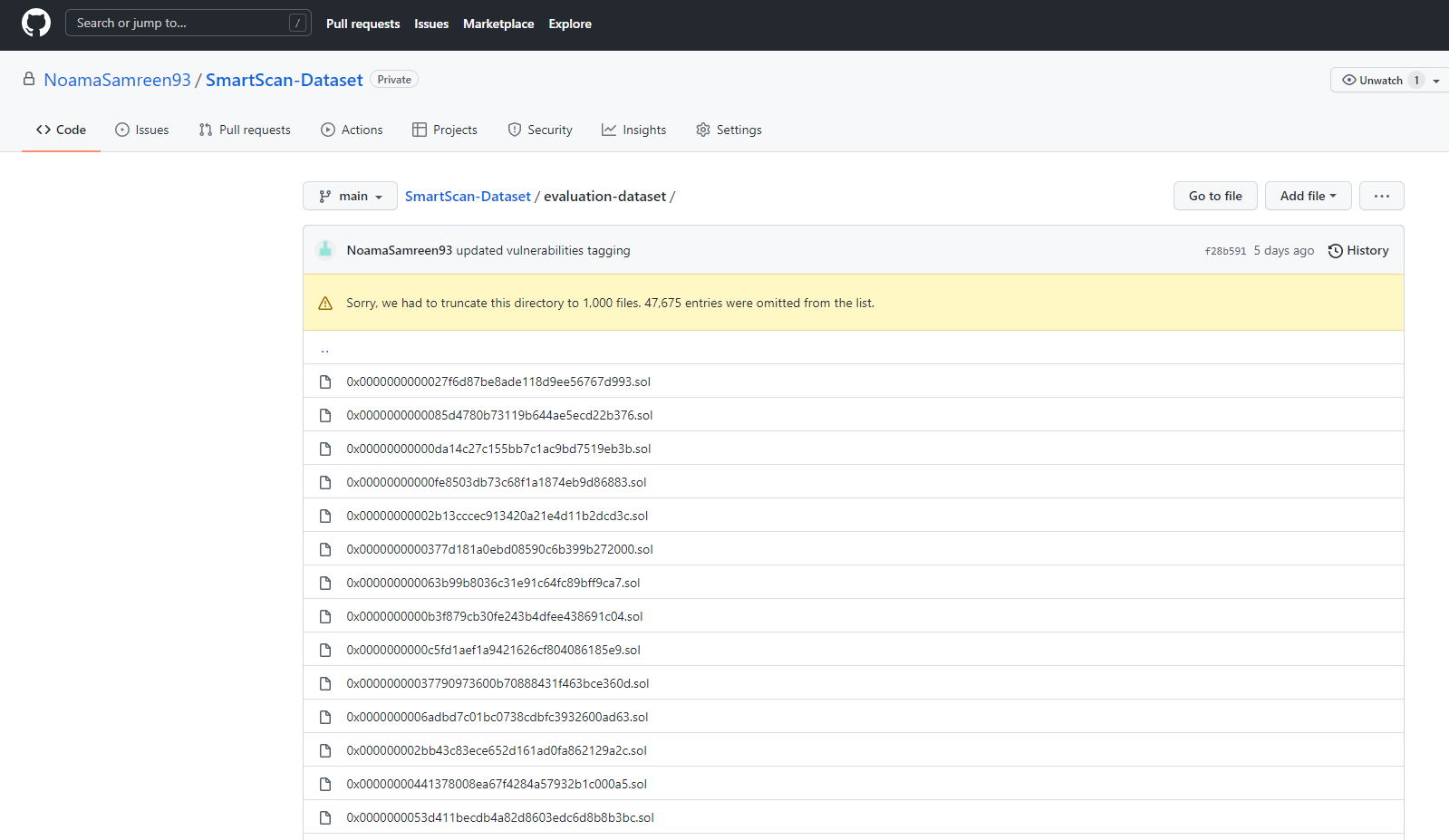}}
\caption{GitHub Repository of used Datasets and Results}
\label{fig:noamaGithub}
\end{figure*}

\begin{figure*}
\centerline{\includegraphics[width=1\textwidth]{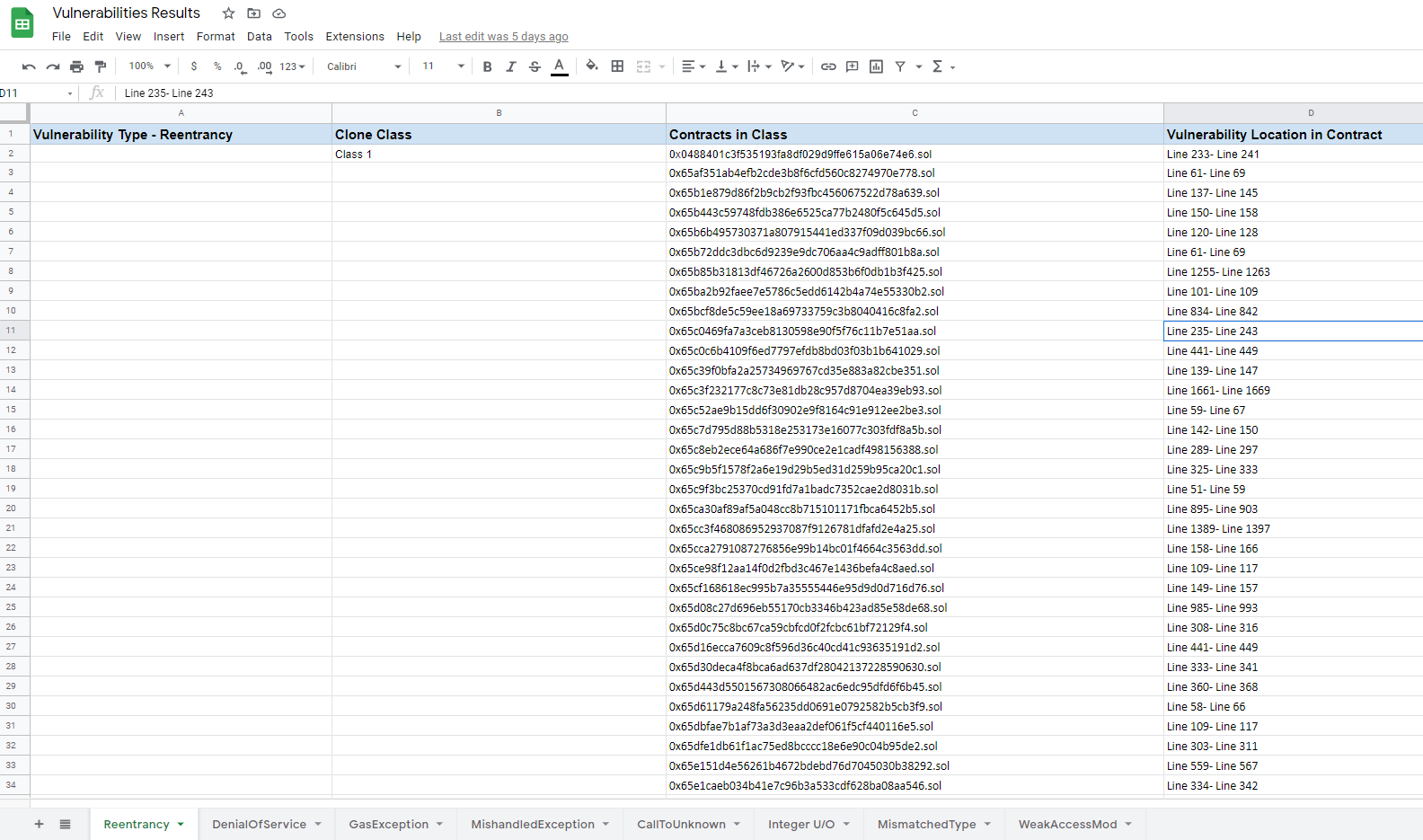}}
\caption{Catalog of Vulnerabilities Detected in Evaluation Dataset}
\label{fig:vulTable}
\end{figure*}

\section{Related Work}\label{relatedwork}
\subsection{Code Clone Detection Approaches for Vulnerability Detection}
Usually, vulnerabilities in a system are detected using static code analysis of the source code. On the other hand, code clone vulnerability in a software system is detected by identifying vulnerabilities that propagate in a system due to code cloning. 
However, our focus of research is to study approaches that supports security vulnerability detection without any adaptations or an approach that uses code clone detection technique within a system to generate a vulnerability code pattern and then use this code pattern to detect its clones and variations in a target system.  

%Table \ref{vulDet} provides a comparison of code clone detection studies from a vulnerability identification perspective. It compares these studies by illustrating their research purposes, detection techniques, and evaluation which is deducted from all the selected approaches. By constructing this table, we observed that most of these articles validated our proposed criteria, which justifies our relevant choice of articles for this review study

Kim et al. \cite{vuddy} addresses the lack of scalability in code clone detection techniques to cover ever growing open-source software systems. This approach targets accurate identification of all types of code clones. It even provides an built-insupport for finding security vulnerabilities in large software programs that is resilient to common modifications in cloned code while preserving the vulnerable conditions. It uses function level granularity and length filtering technique to reduces the number of signature comparisons thereby ensuring a scalable approach. This approach is evaluated using a dataset of C/C++ programs collected from over 25k repositories on GitHub. Vuddy uses vulnerability database such as CVE - Common vulnerabilities and Exposures to discover code clones of vulnerabilities in the target programs. 

Another approach studied in this research that leverages CVE is by Viertel et al. \cite{sourcererCC}. This research work uses an existing token-based code clone detection tool, SourcererCC by Sajnani et al. \cite{Sajnani2021}. This research was selected for study in our review because the authors extend the SourcererCC tool to enable inter-project clone detection to find clones between an external code repository and a project. This adaptation of the SourcererCC matches our proposed criteria of using code cloning for security vulnerability detection. Like the evaluation of Vuddy by Kim et al. \cite{vuddy}, this research was also evaluated using open-source C/C++ programs collected from open-source repositories. Viertel et al. analyzed the CVE database content of 20 reported vulnerabilities and detected their patterns in their C/C++ programs-based dataset. 
The following approaches are coherent with the methodology of vulnerability code-pattern generation from a vulnerability database and its detection in another large dataset to understand its evolution. 

Chen et al. \cite{crossClone} uses Nicad code clone detection tool by Cordy and Roy \cite{nicad} to generate android malware signatures from a dataset of malicious android applications as a first step in their approach. Nicad provides a cross-clone detection plugin to identify code clones in two different systems. Chen et al. leverages this plugin in the second step of their approach to detect malware and its variations in a dataset of benign applications. The authors evaluate their approach by categorizing over 1000 android applications into 19 malware groups and generated malware signature of the 19 malware groups from a vulnerability extraction dataset. They proceed in detecting malware and its variations in a dataset of benign applications with very low false positives and 96.88\% accuracy of malware detection. 

Finally, Liu et al. \cite{vfdetect} proposes a new code clone detection tool, VFDetect, that follows the methodology of vulnerability code pattern generation and identification in a target system. This approach uses vulnerability fingerprinting using hashing of the feature sequences extracted from the diff files in a vulnerability database. This approach validates our review criteria by detecting vulnerabilities in a target system by matching the fingerprint of the vulnerability generated. The authors evaluate their approach by comparing the performance and efficiency of their approach with other code clone detection tools such as, ReDeBug\cite{Jang2012ReDeBugFU} and Vuddy\cite{vuddy}. This is the only research study in our review that compares its results to other available state of the art tools. 

\subsection{Analysis and Discussion of Related Work}
This Section discusses all studied approaches cited in the related work of this research, by following a criterion for inclusion in our review. First, we have identified the criteria representing both the technique used as well as a baseline for a comparison of target systems, purpose, and evaluation of the studied approaches.
The main criteria of the considered approaches are year of publication, code clone detection technique used, whether it is a novel approach, extension to an existing approach or an application of an existing approach, purpose of the research and its evaluation. 

Few research works targeting code cloning in Ethereum smart contracts were not included in the review as they did not justify the defined criteria of approach inclusion. He et al. \cite{characterizing} and Kondo et al. \cite{hassan} are two research studies that investigated code clones in Ethereum smart contracts. However, none of these research studies highlights the security vulnerabilities and their identification using code cloning.  

Based on the used criteria, these approaches can be divided into three types. An built-in category with the support for vulnerability detection incorporated in the proposed approach. A second category where an existing tool is extended to include vulnerability detection by modifying the architecture of the underlying technology used. Lastly, a third category that applies an existing code clone detection tool at different stages of a proposed framework to include vulnerability detection possible. 

For the type of code clones identified in the studied approaches, we observed that most approaches do not explicitly mention it. However, only one work
\cite{vuddy} specifies that their approach covers identification of all the types of code clones. 

In terms of suitability of the studied approaches to Ethereum smart contracts, we conclude that all the approaches studied can be used to detect the vulnerabilities and its evolution in Ethereum smart contracts, however, we notice that \cite{crossClone} uses Nicad \cite{nicad} which is the only code clone detection tools discussed in this study that provides support for Solidity programming language. Another point is that all the studied approaches have selected C/C++ programming language as the target system in their evaluation. 
\begin{table}
\caption{Comparative table of the studied approaches using Code Clone Detection techniques for vulnerability detection}
\label{vulDet}
 \begin{tabular}{|p{2.5cm}|p{4cm}|p{4cm}|p{4cm}|} 
 \hline
 \textbf{Authors/ Year} &  \textbf{Code-Clone Detection Technique} & \textbf{Purpose} &  \textbf{Evaluation}  \\ 
 \hline
 Kim et al./ 2017 \cite{vuddy} & VUDDY- A scalable Approach for Vulnerable Code Clone Discovery  & Increasing the scalability of vulnerability detection in a large software system by reducing the number of signature comparisons & Improved performance and efficiency of detecting vulnerable clones. \\
 \hline
 Viertel et al. / 2019\cite{sourcererCC} & SourcererCC - Detecting Security Vulnerabilities using  Clone Detection and Community Knowledge & Harnessing available example source code of software vulnerabilities, from a large-scale vulnerability database, which are matched to code fragments using clone detection & Analyzed the CVE database content of 20 reported vulnerabilities and detected their patterns in open-source repositories of C/C++ based projects\\
\hline
    Chen et al. / 2015\cite{crossClone} & Nicad - Detecting Android Malware Using Clone Detection& Vulnerability signature generation using code clone detection in a vulnerable database and identifying these vulnerable code signatures and its variations in a generic database of android applications & categorized over 1000 malicious apps in 19 malware families and detected vulnerabilities in a benign applications dataset with very low false positives and high accuracy at 96.88\%. \\
    \hline
    Liu et al. / 2017\cite{vfdetect} & VFDETECT: A Vulnerable Code Clone Detection System Based on Vulnerability Fingerprint & uses vulnerability fingerprinting to match the pre-processed code blocks in target project with the fingerprint & Evaluates by comparing results from other code clone detection tools such as ReDeBug and Vuddy \\
\hline
\end{tabular}
\end{table}
\section{Threats to Validity}
\subsection*{\textbf{Threats to Internal Validity}}
One of the factors that can affect our results is that our approach, VOLCANO, is a signature-based vulnerability detection method which can only detect instances of known vulnerabilities. Though our method can detect variants in vulnerabilities, the extent of variations is limited to the threshold used for clone comparison. The vulnerability detection in our method is based on clone clusters and to form a clone cluster, our detection tool requires at least two similar code fragments. Thus, we have to eliminate some known vulnerabilities from the Vulnerability Dataset as we could not find any clone cluster for those vulnerabilities in the Evaluation Dataset. 
\subsection*{\textbf{Threats to Construct Validity}}
The dataset analyzed in this research consisted of a subset of all the Solidity smart contracts deployed onto the Ethereum main net that corresponds to ÐApps. In this paper we have established an empirical analysis on a dataset that is not equally distributed in terms of various versions of Solidity programming language. The fact that some of the versions of Solidity programming language contained higher number of smart contracts than others may have introduced a measurement bias that can only be overcome with an even dataset that includes reasonably similar smart contracts demographics for each version of Solidity programming language like the total number of smart contracts, their size and complexity. We recognize that further research with an updated dataset may strengthen the vulnerabilities evolution identified in our results.
\subsection*{\textbf{Threats to External Validity}}
The generalizability of our results can be affected by the fact that our dataset contains only a sub-set of smart contracts from the Ethereum blockchain network.  Our Vulnerability Dataset and Evaluation Dataset are not the most up-to-date, and it only contains the smart contracts from January 2020 to December 2020. 

\section{Conclusion and Future Work}
 Our research focuses on investigating the vulnerability classification, identification and evolution efficiently in Ethereum smart contracts. This paper highlights the use of code clone detection techniques in this regard. This study provides a review of the research status on use of code clone detection techniques as a replacement to the usual static analysis for vulnerability identification. While filtering the approaches that use code cloning for generating vulnerability signature from a vulnerability database, we have observed that few studies have been conducted concerning this topic. After analyzing all the studied approaches, we were convinced that Ethereum smart contracts can benefit from code clone detection techniques particularly in vulnerability signature generation and identification of its variations in different versions of Solidity programming language. Therefore, following this literature review, we developed a framework that use code clone detection technique for identifying vulnerabilities and their variations in smart contracts.  We evaluated VOLCANO on two datasets, one with confirmed vulnerabilities and another with approximately 50k random smart contracts collected from the Etherscan\cite{Etherscsan}. This research provides promising results in the field of vulnerability evolution and in the future, we would expand our Vulnerability Dataset to include known vulnerabilities from databases like CVE\cite{CVE} or OWASP\cite{OWASP} to detect potential vulnerabilities in Ethereum smart contracts. 

\bibliographystyle{IEEEtran}
\bibliography{DoctoralSymBib.bib}
\nocite* 
\end{document}